\documentclass[sigconf, nonacm]{acmart}

\newcommand\vldbdoi{XX.XX/XXX.XX}
\newcommand\vldbpages{XXX-XXX}
\newcommand\vldbvolume{14}
\newcommand\vldbissue{1}
\newcommand\vldbyear{2020}
\newcommand\vldbauthors{\authors}
\newcommand\vldbtitle{\shorttitle} 
\newcommand\vldbavailabilityurl{URL_TO_YOUR_ARTIFACTS}
\newcommand\vldbpagestyle{plain}

\usepackage{graphicx}
\usepackage{subcaption}
\usepackage{booktabs} 
\usepackage{placeins} 
\usepackage{dblfloatfix} 
\usepackage{afterpage} 
\usepackage{multirow}
\usepackage{enumitem}

\usepackage{algorithm}
\usepackage{algorithmic}  

\usepackage[utf8]{inputenc}
\usepackage[T1]{fontenc}
\usepackage{minted}          
\usepackage{xcolor}          
\usepackage[most]{tcolorbox} 

\hypersetup{colorlinks=true, urlcolor=blue}

\setminted{
    fontsize=\small,
    breaklines=true,         
    frame=lines,             
    framesep=2mm,
    linenos=true,            
    numbersep=1pt,
    fontfamily=tt
}

\begin{document}
\title{To GPU or Not to GPU: Vector Search in Relational Engines}


\author{Vasilis Mageirakos}
\affiliation{%
  \institution{ETH Zürich, Switzerland}
}
\email{vmageirakos@inf.ethz.ch}

\author{Joel André}
\affiliation{%
  \institution{ETH Zürich, Switzerland}
}
\email{joel.andre@inf.ethz.ch}

\author{Marko Kabić}
\affiliation{%
  \institution{ETH Zürich, Switzerland}
}
\email{marko.kabic@inf.ethz.ch}

\author{Bowen Wu}
\affiliation{%
  \institution{ETH Zürich, Switzerland}
}
\email{bowen.wu@inf.ethz.ch}

\author{Yannis Chronis}
\affiliation{%
  \institution{ETH Zürich, Switzerland}
}
\email{chronis@inf.ethz.ch}

\author{Gustavo Alonso}
\affiliation{%
  \institution{ETH Zürich, Switzerland}
}
\email{alonso@inf.ethz.ch}


\newcommand{\note}[1]{\begingroup\color{blue}\textbf{NOTE: } #1\endgroup}
\newcommand{\todo}[1]{\begingroup\color{red}\textbf{TODO: } #1\endgroup}
\newcommand{\unsure}[1]{\begingroup\color{cyan}#1\endgroup}
\definecolor{darkgreen}{RGB}{0, 150, 0}
\newcommand{\question}[1]{\begingroup\color{darkgreen}\textbf{Q: } #1\endgroup}

\newtcolorbox{insightbox}{
  colback=gray!8, colframe=gray!40,
  boxrule=0.4pt, arc=2pt,
  left=1pt, right=1pt, top=0pt, bottom=0pt,
  before skip=5pt, after skip=5pt
}
\newcommand{\insight}[1]{%
  \begin{insightbox}\textbf{Insight.}\enspace #1\end{insightbox}}
\newcommand{\keyinsight}[1]{%
    \begin{insightbox}\textbf{Key Insight.}\enspace #1\end{insightbox}}



\newpage


\begin{abstract}

Vector search (VS) is now available in most database engines. However, while vector search is a common feature in AI/ML/LLMs where the dominant computing platforms are GPUs, existing database engines operate on CPUs even when implementing vector search. This raises the question of whether integrating vector processing on GPUs as part of the engine would be a better design. In this paper, we explore this question in detail. First, we extend the TPC-H benchmark with vector data (from text and images) and propose a number of representative SQL+VS queries. Second, we develop a modular execution engine that can run SQL+VS queries across CPU and GPU. Third, we perform extensive experiments on a number of deployments: running the SQL+VS queries across CPU and/or GPU, with data residing in CPU or GPU memory, with existing indices and novel, optimized versions, as well as across different GPUs and interconnects (PCIe, NVLink). The results provide actionable and counter-intuitive insights on how to run such queries over CPUs and GPUs. For instance, the relational components benefit much more from running on the GPU than the vector search part. In addition, when the vector search involves moving data and indexes, using the GPU is not the best option, even with fast interconnects. Thus, we develop an alternative organization of vector index and embeddings that reduces the size of the index, making GPU-based vector search more competitive. With these improvements, the final result is that both the relational and vector search components are faster on the GPU, particularly on fast interconnects, in contrast with the architecture used in existing engines.  

\end{abstract}

 \maketitle

\pagestyle{\vldbpagestyle}
\begingroup\small\noindent\raggedright\textbf{PVLDB Reference Format:}\\
\vldbauthors. \vldbtitle. PVLDB, \vldbvolume(\vldbissue): \vldbpages, \vldbyear.\\
\href{https://doi.org/\vldbdoi}{doi:\vldbdoi}
\endgroup
\begingroup
\renewcommand\thefootnote{}\footnote{\noindent
This work is licensed under the Creative Commons BY-NC-ND 4.0 International License. Visit \url{https://creativecommons.org/licenses/by-nc-nd/4.0/} to view a copy of this license. For any use beyond those covered by this license, obtain permission by emailing \href{mailto:info@vldb.org}{info@vldb.org}. Copyright is held by the owner/author(s). Publication rights licensed to the VLDB Endowment. \\
\raggedright Proceedings of the VLDB Endowment, Vol. \vldbvolume, No. \vldbissue\ %
ISSN 2150-8097. \\
\href{https://doi.org/\vldbdoi}{doi:\vldbdoi} \\
}\addtocounter{footnote}{-1}\endgroup

\ifdefempty{\vldbavailabilityurl}{}{
\vspace{.3cm}
\begingroup\small\noindent\raggedright\textbf{PVLDB Artifact Availability:}\\
The source code, data, and/or other artifacts have been made available at \url{https://github.com/mageirakos/vec-h}.
\endgroup
}

\section{Introduction}
\label{sec:intro}
Vector databases offer nearest-neighbor and filtered vector search capabilities, many with GPU-accelerated index construction and query serving~\cite{wang2021milvus, weaviate2025, pinecone2025}.
These capabilities are being subsumed by relational database systems~\cite{stonebraker2024goes} by integrating vector search (VS) as a standard operator, which has become a common feature in most engines~ \cite{pgvector2025, oracle_aivs2025, scann_alloydb2025, li2025gaussdb, chen2024singlestore, wei2020analyticdb, upreti2025cost}. This integration enables vector search over embeddings to be combined with the expressiveness of SQL. Vector search in relational database systems runs on CPUs, since existing engines typically do not use GPUs. This difference in hardware support is important to study, as relational workloads can run significantly faster on GPUs~\cite{maximus_marko,yogatama2026sirius,wu2025terabytescaleanalyticsblinkeye,mohr23-boss,wu25-gpu-joins-groupby}, by using modern interconnects such as NVLink~\cite{maxbench_marko,lutz2020pump}, while vector search itself is already routinely run on GPUs in ML systems~\cite{wang2021milvus,johnson2021faiss}.

Vector search (VS) engines, such as FAISS~\cite{douze2025faiss} and cuVS~\cite{cuvs}, 
identify data movement as the primary bottleneck: transferring embeddings and index structures across the PCIe bus accounts for the majority of end-to-end runtime, exceeding 90\% of query time for large indexes~\cite{liu2026gpu_survey}. To mitigate this, VS engines either keep index data resident on the GPU~\cite{ootomo2024cagra, douze2025faiss, xi2025vecflow}, overlap data transfers with computation~\cite{karthik2024bang}, or explore hybrid CPU-GPU tiering~\cite{zhang2024rummy}. When it comes to vector search, however, database (DB) engines take a different approach: several DB engines use GPUs to accelerate index building, but execute queries on the CPU~\cite{oracle_aivs2025,scann_alloydb2025}. Other DB engines do not use GPUs at all in either index creation or query execution~\cite{pgvector2025,duckdb_vss2024}. There are several reasons for this. The embedding data is typically orders of magnitude larger than relational data. With 1024-dimensional embeddings, vector data can be one order of magnitude or more larger than the relational dataset. This means keeping full indexes on GPU memory is impractical since that memory is smaller, more expensive, and shared with other analytical operators. Additionally, research on VS engines confirms that interconnect bandwidth limits GPU querying, and that large batch sizes are needed to amortize transfer costs~\cite{johnson2021faiss}. Database workloads rarely reach those batch sizes and, typically, do not operate on batches of queries. These concerns are reinforced by how database engines currently use VS indexes: the index stores the actual embeddings, moving an index involves moving all the embeddings.

In this paper, we explore the use of heterogeneous CPU-GPU architectures in the context of running analytical SQL+VS queries combining relational and vector search operators. We aim to empirically answer three central questions: a) \textit{What are the performance characteristics and requirements of analytical SQL+VS queries} b) \textit{Does using a GPU accelerate such queries}, and c) \textit {How should the vector indices and query execution be designed to maximize performance?}
These are non-trivial questions, as they involve elements that interact with each other in non-intuitive ways: architectural regarding the interconnects, engine design regarding where the queries run, and data structure design regarding the way the indexes are constructed. 

To answer these questions, we propose a novel analytical vector search benchmark and develop a system to seamlessly run them across CPU-GPU. The benchmark, called \textbf{Vec-H}, extends TPC-H with two new tables of real semantic embeddings and eight representative analytical SQL+VS queries spanning five VS integration patterns (inner, left, lateral, anti, semi). The queries explored cannot be expressed in the query languages of pure vector databases (i.e., no SQL support)~\cite{wang2021milvus, weaviate2025, pinecone2025}. For the system, we developed \textbf{MaxVec}, a CPU-GPU execution engine for SQL+VS queries. It extends Maximus~\cite{maximus_marko}, an accelerated CPU-GPU query engine for data analytics, with VS operators for exhaustive nearest-neighbor search (ENN) and approximate nearest-neighbor search (ANN) via FAISS/cuVS \cite{douze2025faiss,cuvs}.
Each operator can be placed on a CPU or a GPU independently, and the engine handles index and data movement across devices transparently. To our knowledge, MaxVec is the only open-source engine that supports full SQL+VS on a hybrid CPU-GPU platform.

Based on this infrastructure, we perform extensive experiments using existing vector search libraries and algorithms, all of which are available for extension and further experimentation. We run Vec-H using MaxVec on three hardware configurations with PCIe~5.0, NVLink-C2C, and unified CPU-GPU memory (DGX Spark \cite{nvidia_dgx_spark}). The results are a series of important and counterintuitive insights:

Regarding CPU vs. GPU execution, we have experimentally tested where to run queries, entirely on the CPU, the GPU, or hybrid. This initial analysis leads to two surprising insights. First, using standard vector search libraries and indexes, GPU execution yields a bigger improvement for relational operators than for vector search. Second, relational operators can be orders of magnitude more expensive than a vector search. 

Regarding faster interconnects, MaxBench \cite{maxbench_marko} shows that for relational analytics, data transfers account for 67\textendash 98\% of runtime on PCIe, but on NVLink the interconnect bottleneck disappears and execution becomes GPU bound. Our experiments show that this is not the case for vector search. Even when using NVLink, 
with current data-owning vector indexes, running vector search on a GPU does not pay off unless the vector index is resident on the GPU. This is because the size of the index is so large, and its transfer incurs per-call setup and CPU-to-GPU index transformation overheads, that the increase in bandwidth is not enough to compensate for the added latency of having the data movement.

In terms of the index itself, we show that by separating the actual index (comparatively very small) from the storage of the embeddings (very large), it is possible to move the index to the GPU when needed, making running both the relational and vector search components on the GPU the faster option even when the data for the relational and vector search have to be transferred to the GPU.

\section{Related Work}
\label{sec:related_work}
\noindent\textbf{GPU vector search.}
GPU-accelerated VS libraries achieve speedups over CPU implementations for both index construction and query execution~\cite{ootomo2024cagra, karthik2024bang, zhang2024rummy, liu2026gpu_survey}.
FAISS~\cite{douze2025faiss} is widely used, supporting CPU and GPU execution on multiple index types.
Its GPU backend integrates cuVS, which provides CAGRA~\cite{ootomo2024cagra} and other GPU-optimized indexes. 
CAGRA, the current GPU graph-based index SOTA, reports 33--77$\times$ higher throughput than CPU HNSW at 90--95\% recall.
For datasets that exceed GPU memory, BANG~\cite{karthik2024bang} and RUMMY~\cite{zhang2024rummy} split data between the GPU and the host. The tradeoff is higher search latency from fetching data over the interconnect during search.
VecFlow~\cite{xi2025vecflow} is the first purpose-built GPU index for filtered search, but supports only categorical label predicates, not arbitrary SQL predicates.

\noindent\textbf{Database-integrated vector search.}
Vector databases such as Milvus~\cite{wang2021milvus} support GPU-accelerated VS and filtered vector search (FVS)~\cite{fvs_chronis_survey_vldb}, although filters run on the CPU.
They focus on VS and FVS workloads rather than general queries.
Several database systems integrate VS with CPU-only query execution~\cite{pgvector2025, zhang2023vbase, wei2020analyticdb, chen2024singlestore, oracle_aivs2025, scann_alloydb2025, li2025gaussdb}.
The two closest GPU SQL engines, Theseus~\cite{aramburu2025theseus} and Sirius DB~\cite{yogatama2026sirius}, both list vector search as future work.
Our work fills this gap with a query engine capable of full SQL+VS (ENN and ANN) on heterogeneous CPU+GPU hardware.

\noindent\textbf{Interconnect and data movement.}
Lutz et al.~\cite{lutz2020pump} provide the foundational NVLink-vs-PCIe study for database workloads. More recently, MaxBench~\cite{maxbench_marko} extends this to modern hardware (PCIe~5.0, NVLink-C2C) and shows that data transfers account for 67--98\% of relational query runtime on PCIe, while NVLink shifts execution to be GPU-bound.
A recent GPU VS survey~\cite{liu2026gpu_survey} confirms the pattern for vector search on PCIe, where data transfers consume 40-97\% of the total latency, but does not extend the analysis to modern hardware.
VecFlow-Chamfer~\cite{mo2026vecflowchamfer} uses cache-coherent host-memory access over NVLink-C2C, a hardware capability of NVIDIA's Grace Hopper Superchip~\cite{nvidia_grace_hopper}, to read from host memory during the final re-ranking step of their pipeline. 
We study and utilize these hardware capabilities for different workloads and indexes, and characterize which design wins and when, in Section~\ref{sec:experiments}.

\noindent\textbf{Related benchmarks.}
Existing benchmarks~\cite{aumuller2020ann, zhu2025experimental, bigvectorbench_vldb} evaluate standalone nearest-neighbor search or add simple predicate evaluation on embedding metadata.
On the database analytics side, TPC-H and TPC-DS~\cite{tpch, tpcds_spec} are the standard relational benchmarks but do not contain vector search operators. The unique advantage of relational databases with integrated VS support is access to the full relational data model and the expressiveness of SQL. Under existing benchmarks, it is difficult to differentiate database systems from specialized vector databases. The closest to our benchmark are SemBench~\cite{lao2025sembench} and Exqutor~\cite{kim2026exqutor}. SemBench combines SQL with semantic operators, but uses LLM invocations whose ``processing overheads dominate total computation costs''~\cite{lao2025sembench}. Exqutor targets cardinality estimation for vector predicates on CPU, attaching embedding columns to TPC-H and TPC-DS for a range-based distance filter over the original queries. We instead target execution on heterogeneous CPU/GPU hardware, with Vec-H queries that span multiple VS integration patterns (inner, left, lateral, anti, semi), real semantic embeddings, and two new tables joined to TPC-H.

\section{Vec-H Benchmark}
\label{sec:benchmark}
We design the Vec-H benchmark to study and optimize SQL+VS queries. It targets analytical SQL+VS queries that combine nearest-neighbor search with joins, aggregates, and ordering. We extend TPC-H with two tables containing embeddings and eight unique query templates.
We treat exact embedding models, dimensionality, and data source as adjustable parameters.

\subsection{Dataset}
\label{sec:vech-dataset}

\begin{figure}[t]
  \centering
  \includegraphics[width=.95\linewidth]{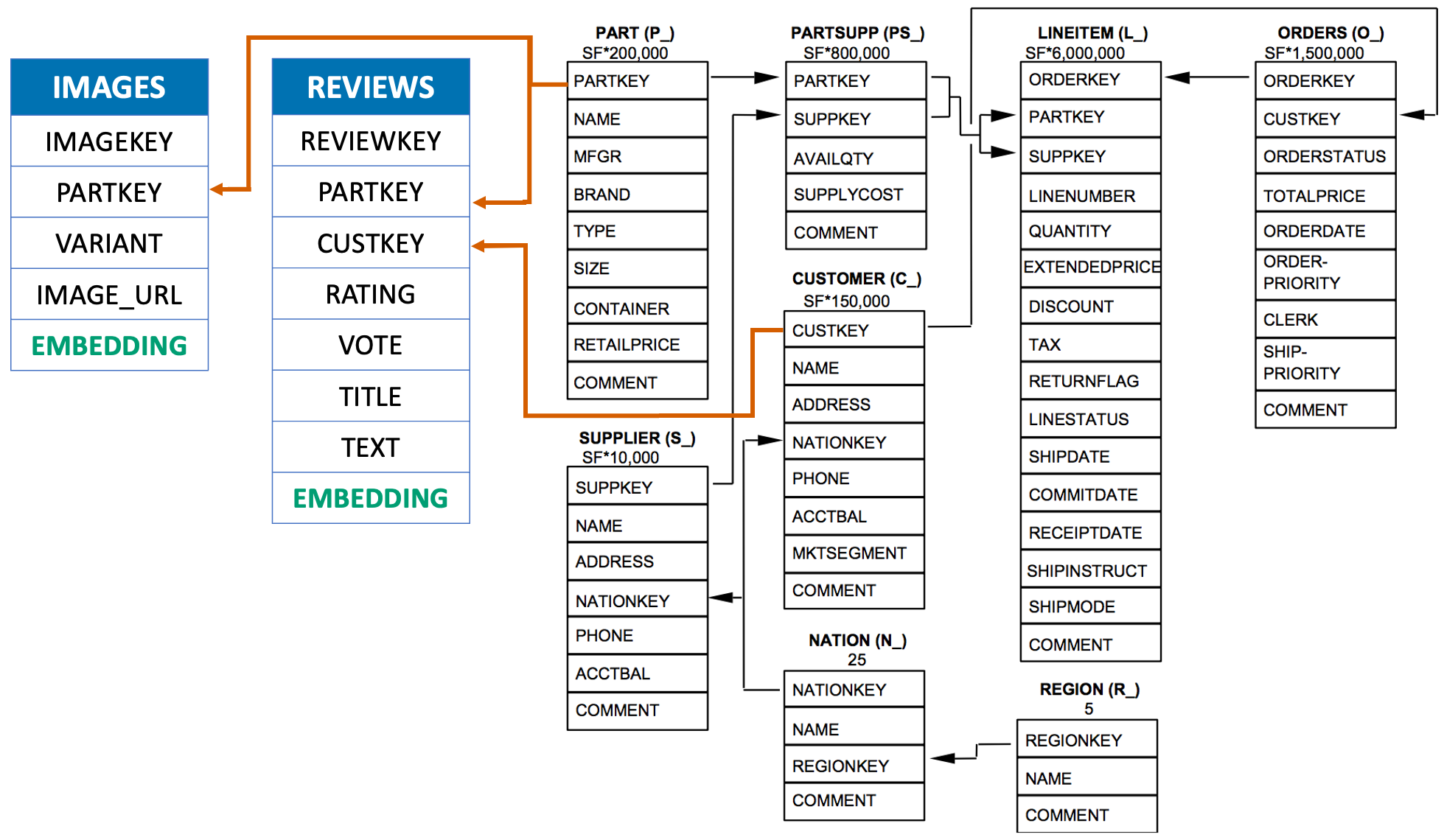}
  \caption{Vec-H schema: TPC-H extended with IMAGES and REVIEWS tables, each with an EMBEDDING column.}
  \label{fig:vech-schema}
\end{figure}

\noindent\textbf{Data source.}
Vec-H adds two tables to TPC-H, the REVIEWS and IMAGES, populated with data from the Amazon Reviews dataset~\cite{hou2024bridginglanguageitemsretrieval}, which covers 34 product categories of 48M unique products with 571M reviews, and 133M images.

\noindent\textbf{Schema and mapping.}
The extended schema is shown in Figure~\ref{fig:vech-schema}. We establish a one-to-one mapping between TPC-H parts and Amazon products: each \texttt{P\_PARTKEY} is randomly assigned a single \texttt{parent\_asin} (Amazon's unique product identifier). All reviews and images of the sampled product are then added to the reviews and images tables and linked to their part via foreign keys, minus those dropped during cleaning (e.g., broken image URLs). Each review is also assigned to a customer by mapping the Amazon \texttt{user\_id} to a \texttt{C\_CUSTKEY}. If distinct Amazon users exceed the fixed TPC-H customer count for a given scale factor (SF), we assign the reviews randomly to existing \texttt{C\_CUSTKEY}s. Thus, we preserve the TPC-H customer cardinality per SF and the part-to-review real world distribution, but not the original distribution of reviews across Amazon users. Once the mapping is fixed, \texttt{P\_PARTKEY} acts as the product identifier: REVIEWS and IMAGES join to PART and CUSTOMER.
Reviews and images are provided by the Amazon Reviews dataset in their raw form (text, image URLs). We generate one embedding per review and per image using open-source models.

\noindent\textbf{Dataset scale.}
We use the SF-based scaling of TPC-H.
At SF=1, the TPC-H dataset is ${\sim}$1\,GB.
For the embeddings, at a given SF, the total data size depends on embedding dimensionality and bytes per dimension. The number of reviews and images is well-approximated as $\text{SF} \times 200{,}000 \times \overline{R}$ and $\text{SF} \times 200{,}000 \times \overline{I}$ respectively, where $\overline{R} {\approx} 12$ and $\overline{I} {\approx} 4$ are the average reviews and images per product across all product categories. Products may have reviews and no images or vice versa. Review counts per product are long-tailed, while image counts are more normally distributed.

Since review and image embeddings may originate from models with different dimensions $d_r$ and $d_i$, the total embedding data size is approximated by:
$\text{Vec\_bytes} \approx SF \times 200{,}000 \times (\overline{R} \times d_r + \overline{I} \times d_i) \times b$, where $b$ are the bytes used per dimension. The Rel:VS  data ratio is the TPC-H data size ($\sim\!$1\,GB per SF) to the embedding data size ($\text{Vec\_bytes}$) in gigabytes. It is dependent on the embedding dimensions $d_r$ and $d_i$, the bytes per dimension $b$, and the product-to-part mapping, which determines $\overline{R}$ and $\overline{I}$ per part. The Amazon Reviews dataset~\cite{hou2024bridginglanguageitemsretrieval} has 48M unique products, and thus enough data to support a Vec-H with {SF${\approx}$240} (48M parts), after which all 571M reviews and 133M images have been assigned. The scale factor, embedding models, dataset configuration, and Rel:VS we used are listed in Section~\ref{sec:experiments}.

\subsection{Logical Vector Search Operator}
\label{sec:logical_vs_operator}
We define a vector search operator logically as \textit{vector\_search(query vectors, embedding column, k)}, where for every query vector the operator finds the $k$ closest embeddings. We allow for two inputs in the operator (query vector and embedding data sides). This way, the \textit{vector\_search} operator is binary. It can be used as a similarity join operator~\cite{silva_similarity_join_icde, fast_annjoin_paper}, operating on a batch of queries, or a simple vector search when the \textit{query vector} input side has a cardinality of 1. 
We discuss the design in more detail in Section~\ref{sec:maxvec_vector_operators}.

\begin{figure}[t]
  \centering
  \includegraphics[width=\linewidth]{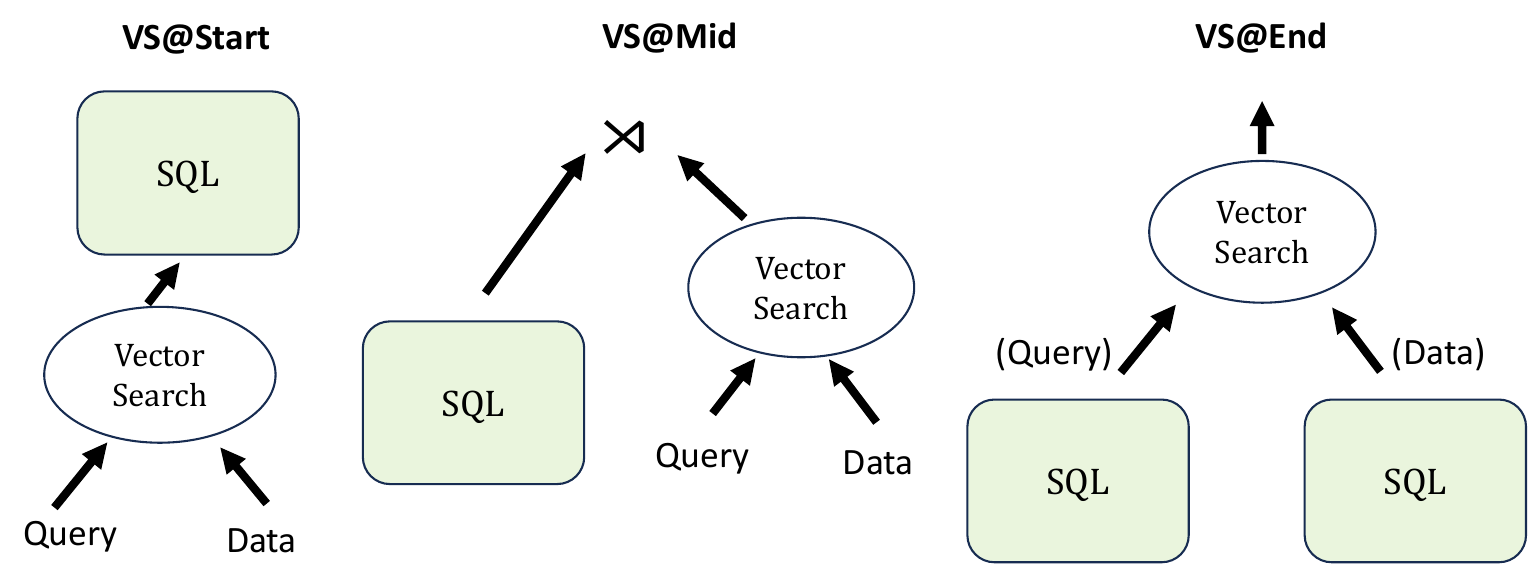}
  \caption{Vec-H query reference implementations in MaxVec.}
  \label{fig:vech-query-types}
\end{figure}

\begin{table*}[t]
    \centering
    \caption{Vec-H query summary. Each query extends a TPC-H base with vector search. Join type describes how VS results integrate with the relational plan.}
    \label{tab:vech-queries}
    \small
    \begin{tabular}{@{}clllp{5.5cm}@{}}
        \toprule
        \textbf{Query} & \textbf{TPC-H Base} & \textbf{VS Use Case} & \textbf{Join Type} & \textbf{Key Distinction} \\
        \midrule
        \multicolumn{5}{@{}l}{\textit{VS@Start}} \\
        Q2  & Min-cost supplier & Image similarity filter & Inner join & Join order sensitivity to $k$, distance in \texttt{ORDER BY} \\
        Q16 & Parts/supplier count & Semantic review exclusion & Anti-join & Exclusion changes \texttt{GROUP BY} aggregate counts \\
        Q19 & Discounted revenue & Dual VS OR branches & Semi-join, $\times$2 & Two vector searches on different modalities \\
        \midrule
        \multicolumn{5}{@{}l}{\textit{VS@Mid}} \\
        Q10 & Returned item revenue & Review similarity flag & Left join & VS annotates matches \\
        Q13 & Customer distribution & Review match count & Left join & VS adds new \texttt{GROUP BY} aggregation dimension \\
        Q18 & Large volume customer & Visual quantity score & Left join & VS contributes numeric aggregate \\
        \midrule
        \multicolumn{5}{@{}l}{\textit{VS@End}} \\
        Q11 & Important stock & Visual duplicate detection & Left join lateral & Forced execution order, batched per-row VS \\
        Q15 & Top supplier & Semantic review ranking & Inner join & SQL scopes VS data, symmetric to VS@Start \\
        \bottomrule
    \end{tabular}
    \par
\end{table*}

\subsection{Queries}
\label{sec:vech-queries}

Each Vec-H query directly extends its TPC-H counterpart (Vec-H Q2 extends TPC-H Q2, and so on). Readers familiar with TPC-H can identify the core relational backbone of the query. To connect the VS result to the relational plan, every query introduces at least one additional join. Table~\ref{tab:vech-queries} summarizes the set of queries, their join types, and the aspect of analytical SQL+VS execution targeted.
We classify queries by where the VS operator appears in the execution plan: VS@Start, VS@Mid, or VS@End (Figure~\ref{fig:vech-query-types}).
These labels reflect data dependencies between the VS and relational subplans. Below we describe the Vec-H queries along with an example business logic of use cases they support. Full SQL definitions and sample result outputs are available in the artifact repository.

\subsubsection{{VS@Start Queries}}

\label{sec:vech-start}

In these queries, vector search is the \textit{query driver}. On the VS operator, the data side is a base table, and the query side is a user-provided query embedding (Figure~\ref{fig:vech-query-types}). The top-$k$ result set affects downstream operators of the plan, either by inclusion (inner or semi-join) or exclusion (anti-join).

\textbf{Q2: Minimum Cost Supplier.}
TPC-H Q2 finds the minimum-cost supplier in a given region for parts of a specific type and size. Vec-H replaces the part type and size predicate with a top-$k$ image similarity search, selecting the parts whose image is closest to the user query image. The VS returns $k$ images each belonging to a unique part. This result is inner-joined to the rest of the plan on \texttt{p\_partkey}, with $k$ directly controlling the selectivity, acting as an inclusive filter where a larger $k$ has more qualifying parts. For a very large $k$, an optimizer could theoretically choose to reorder the joins, and run part of the relational subquery first. Q2 also uses the VS distance as a secondary sort key in the \texttt{ORDER BY}. \noindent\textit{Business logic:} Given a query image, find the suppliers which offer the minimum cost for the $k$ most visually similar parts.

\textbf{Q16: Parts/Supplier Relationship.}
The TPC-H Q16 query excludes suppliers whose comment field contains ``Customer'' and ``Complaints''. We substitute the substring matching predicate with vector search, to exclude suppliers associated with the $k$ most similar reviews to a query embedding via \texttt{NOT IN}. Unlike Q2's inclusive filter, Q16 affects downstream operators in the opposite way.
\noindent\textit{Business logic:} What is the number of ``trustworthy'' suppliers for a given part group (brand, type, size). Trustworthiness is defined by excluding suppliers linked to parts with reviews related to a specific complaint, a negative keyword, or a topic of concern.

\textbf{Q19: Discounted Revenue.}
TPC-H Q19 computes the discounted revenue across three part categories, combined via \texttt{OR}. Vec-H extends the query with two additional VS-defined categories, one matching parts by review similarity and the other by image similarity. Because the branches are combined with \texttt{OR}, the VS results expand the qualifying set. Q19 is the only Vec-H query requiring two simultaneous VS searches across different indices.\noindent\textit{Business logic:} What is the total discounted revenue for parts identified by a traditional brand/container criteria \texttt{OR} a semantic review description \texttt{OR} a visual reference image.

\subsubsection{{VS@Mid Queries}}
\label{sec:vech-mid}

In these queries, VS and the SQL subplans are in separate query plan branches. VS can enrich the SQL output via \texttt{LEFT JOIN} but does not filter any rows.

\textbf{Q10: Returned Item Reporting.}
TPC-H Q10 ranks the top-20 customers by lost revenue from line items those customers returned within a given quarter. Vec-H adds an independent VS branch that retrieves the global top-$k$ reviews most similar to a query embedding, with no restriction to returned products or to the Q10 customers. The \texttt{LEFT JOIN} on \texttt{c\_custkey} annotates each Q10 output row with an \texttt{is\_in\_top\_k} flag, marking whether that customer happens to appear among the top-$k$ reviewers.
\noindent\textit{Business logic:} Among customers responsible for the most returned-item revenue loss, which ones also authored a review similar to a target complaint? 

\textbf{Q13: Customer Distribution.}
TPC-H Q13 computes the distribution of customers by their number of orders, including customers who have no order record. Vec-H independently retrieves the global top-$k$ reviews by similarity and \texttt{LEFT JOIN}s them to count how many top-$k$ reviews belong to each customer. Unlike Q10 and Q18, where VS annotates the final output rows, Q13's \texttt{LEFT JOIN} sits inside the nested subquery, so the VS-derived count propagates through both \texttt{GROUP BY} levels into the output as a second distribution dimension alongside the original order-count distribution.
\noindent\textit{Business logic:} 
Which customer buckets, grouped by order count, concentrate reviews that are most similar to a query embedding? For instance, reviews similar to complaints made by customers with zero orders may signal fraudulent activity.

\textbf{Q18: Large Volume Customer.}
TPC-H Q18 finds the top-100 customers with the largest-quantity orders. Vec-H \texttt{LEFT JOIN}s the top-$k$ visually similar part images to each line item.
A \texttt{CASE} expression sums \texttt{l\_quantity} only for line items whose part matches the VS result, producing a \texttt{similar\_qty} score per order.
The output is re-ranked by \texttt{similar\_qty}, surfacing orders that contain the most items visually similar to the user query.
Combined with TPC-H Q18's multi-join relational plan and \texttt{HAVING} subquery, this tests whether a system can handle a heavy relational workload alongside a full VS branch.
\noindent\textit{Business logic:} Among customers with the largest orders, how many of their ordered items visually resemble a reference part? Re-rank by this visual match quantity.

\subsubsection{{VS@End Queries}}
\label{sec:vech-end}

In these queries, SQL output feeds VS operator inputs (Figure~\ref{fig:vech-query-types}).

\textbf{Q11: Important Stock Identification.}
TPC-H Q11 identifies parts whose total stock value from a given nation exceeds a fraction of the global value. Vec-H uses each high-value part's image as a per-row query vector to find its nearest visual duplicate via \texttt{LEFT JOIN LATERAL}, excluding self-matches.
Here, the query vectors come from the data itself: each part's own image is used to search for visually similar parts.
An optional $k$ on the CTE can limit how many high-value parts enter the VS stage, controlling the batch size. We set $k$ so that all qualifying rows pass through.
Query vectors do not exist until the TPC-H subquery completes, so the SQL plan must run before any VS work can begin.
\noindent\textit{Business logic:} Among the highest-value stock parts from a given nation, which have visually near-identical counterparts? 

\textbf{Q15: Top Supplier.}
TPC-H Q15 finds the supplier with the highest quarterly revenue. Vec-H then joins through \texttt{partsupp} and \texttt{part} to collect all reviews for that supplier's parts, and ranks them by semantic similarity to a query embedding.
The \texttt{INNER JOIN} restricts which reviews the VS operator searches, scoping the data side rather than the query side.
This is symmetric to VS@Start from the opposite direction: where Q2 uses VS results to filter SQL, Q15 uses SQL joins to restrict VS data.
\noindent\textit{Business logic:} For the top-revenue supplier in a given quarter, what are the most relevant reviews (by semantic similarity to a specific concern) for the parts they supply?

\subsubsection{{Result Quality}}
\label{sec:vech-quality}

Vector search is inherently an approximated operation. Errors, whether originating from the embedding model, the choice of ANN over ENN, or desired value of $k$, can propagate through the query plan and affect the final result.
The core metric we use for result quality assessment is \textit{recall}: the fraction of true output rows that appear in the ANN result. We measure recall at the \textit{query output} level, not at the VS operator level. We treat the result of a query that uses ENN (exhaustive) operator as the ground truth. Q19 requires a query-specific metric, its output is a single \texttt{SUM(revenue)} scalar, so row-level recall collapses to a binary match (100\% if the number is exact, 0\% otherwise). Therefore we report \textit{relative revenue error}, $\mathit{rel\_err} = |\mathit{rev}_{\text{ANN}} - \mathit{rev}_{\text{ENN}}| / \mathit{rev}_{\text{ENN}}$, where $\mathit{rev}_{\text{ANN}}$ and $\mathit{rev}_{\text{ENN}}$ are the aggregate revenue produced by the ANN and ENN runs. Lower error is better, and the metric is scale-free and weights each missed top-$k$ partkey by the revenue it would have contributed through the downstream \texttt{lineitem} join. We target $\leq$1\% relative error for Q19, and at least $95\%$ recall per-query.

Result quality also depends on how $k$ interacts with the relational plan. In Q2, Q15, Q18, and Q19 on ANN we have a post-filter execution pattern where a predicate may discard VS results. Thus, we over-sample top-$k'$, where $k'$ ensures returning $k$ passing rows. On the GPU, the indices in FAISS enforce an upper bound on $k'$ which the CPU path does not have~\cite{johnson2021faiss, faiss_gpu_k_cap}, so if the over-sampling required to achieve a recall target exceeds this limit, the query must fall back to CPU. Q15 is one such case in Vec-H, where the GPU top-k limit is reached, which is noted in Section~\ref{sec:experiments}.

\section{MaxVec Engine}
\label{sec:engine}
MaxVec extends Maximus~\cite{maximus_marko}, a modular query execution engine, to support vector search.
The integration incorporates FAISS \cite{douze2025faiss} with both its CPU and GPU (cuVS-enabled \cite{cuvs}) backends as a new execution engine for vector search. This allows relational and vector search operators to be combined within a single query plan with automatic data movement across devices.
This section describes the extensions required: new data formats for embeddings, index management, and vector search operators.

\subsection{Maximus Background}

Maximus \cite{maximus_marko} is a query execution engine for heterogeneous environments, originally designed for relational data analytics. Its central idea is \emph{operator-level integration}, where different execution engines, potentially targeting different hardware such as CPUs, GPUs and FPGAs, can be combined within a single query plan.
Maximus allows each operator to be independently assigned to a specific device and backend. This is achieved through a namespace nomenclature of the form \texttt{device::engine::operator}. For example, a filter operation can be executed on the CPU using the Acero engine (\texttt{cpu::acero::filter}), while a hash join in the same plan runs on the GPU using cuDF (\texttt{gpu::cudf::hash\_join}).
A key feature of Maximus is its ability to automatically orchestrate data movement and format conversion between operators. On CPU, data is represented in Arrow columnar format and executed by the Acero engine. On GPU, data is represented in cuDF format, which also serves as the execution engine for relational operators.
Conversions between these formats are implemented lazily and are zero-copy whenever possible. Data transfers across devices (e.g., CPU to GPU) are handled transparently and executed asynchronously to minimize overhead. While this design targets relational processing, extending it to vector similarity search introduces new challenges. Vector search requires handling high-dimensional embeddings, specialized index structures, extending memory pool management, defining the semantics of new vector search operators, integrating compute-intensive kernels, and extending the  benchmarking layer to profile index movement and VS operators. Incorporating these elements into the existing operator-level abstraction required substantial extensions across the Maximus architecture.

\subsection{Data Formats and Index Management}

To support vector embeddings within relational tables, we introduce a new column type, \texttt{embeddings\_type}. This type is implemented as a nested list structure, where each tuple contains a vector. On the CPU, this is represented using \texttt{arrow::ListType}, while on the GPU it is implemented using \texttt{cudf::ListDType}. Internally, both representations store all embedding values in a contiguous memory region, enabling efficient access patterns.
This design allows us to provide zero-copy interoperability with FAISS by directly exposing the underlying memory buffers as embedding arrays. As a result, embeddings stored in tables can be used for index construction and query evaluation without additional serialization or copying overhead.
In addition to native embedding column support, FAISS index structures are integrated as transferable objects in the engine, similar to tables. Indices can reside in CPU or GPU memory.
Index transfers use FAISS's built-in functions. We introduce optimizations (memory pinning, index caching and secondary non-data-owning VS indices with host-data access) to reduce transfer overhead.
On the GPU side, we integrate FAISS into the Maximus RMM (RAPIDS Memory Manager) memory pool \cite{rmm}, so that relational data and vector indices are managed under one memory budget with shared CUDA streams and access to pinned memory.

\subsection{Physical Vector Search Operators}
\label{sec:maxvec_vector_operators}

Vector search operators follow the same namespace abstraction, and operate on embeddings in the \texttt{embeddings\_type} format. Database engines such as DuckDB~\cite{raasveldt2019duckdb} allow physical operators like hash joins to expose multiple input ports, each of which can be blocking or streaming. We adopt this design in our vector search operators, which have two input ports, one for the query vector side and one for the data side. The data port is always blocking: the full data side must be materialized before search begins, since queries need to identify their nearest neighbors from the entire data side input. Our operators can operate on a single query or a query batch. On CPU, the query port is streaming: query batches are processed as they arrive. On GPU, the query port is blocking. All query-side vectors are accumulated into a single batch so the distance computation runs as one large matrix multiplication rather than many smaller independent vector products. Distance computation between a $N$-vector query batch and $M$ data vectors of dimension $d$ expands to a $N \times d \times M$ GEMM, so larger batches produce wider GEMM kernels and better utilize GPU compute~\cite{nvidia_matmul_perf}, which is why batching is necessary for higher GPU throughput and utilization for VS.

Before execution, data tables and indices can each reside on CPU or GPU. Data movement is handled at execution time, transferring all inputs to the operator's scheduled target device. The FAISS vector search kernels return raw arrays of the query nearest-neighbor indices and distances. The operator constructs the output table by gathering the corresponding rows from both inputs (query vectors and data) and adding a distance column. Any input column can be projected away on the output, to avoid materializing columns the plan does not use. The same operator serves both single-query vector search and similarity join~\cite{fast_annjoin_paper}. 

\begin{figure}[t]
  \centering
  \includegraphics[width=\linewidth]{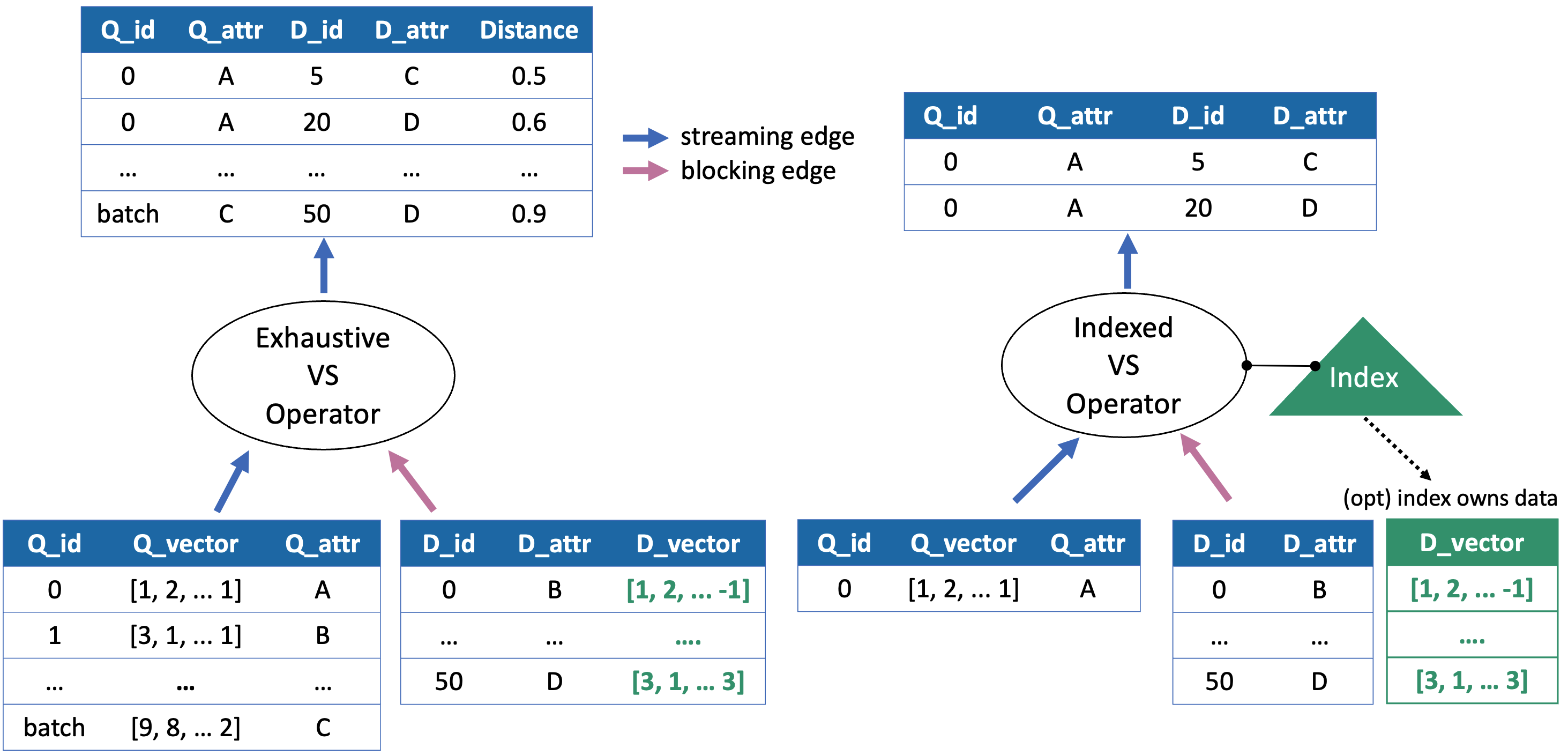}
  \caption{Vector search operators in MaxVec. Exhaustive (left) and indexed (right, where index may own the data). 
  }
  \label{fig:operators}
\end{figure}

Figure~\ref{fig:operators} illustrates an exhaustive VS operator (left) on a batch of queries, and an indexed VS operator (right) that operates on a single query vector. For each $Q\_id$, the output contains the $k=2$ nearest $D\_id$ neighbors and, optionally, their $Distance$, along with other columns from both input sides. When the query side comes from a similarity join (e.g., Vec-H Q11's \texttt{LEFT JOIN LATERAL}), the batch is produced from the join's outer relation.

\subsubsection{Exhaustive VS Operator}
\label{sec:exhaustive-operator}

The exhaustive operator computes distances between every query embedding vector and every data vector from the data-side table, i.e., it is an exact nearest neighbor search (100\% recall). It uses FAISS's brute-force search kernels on the \texttt{embeddings\_type} columns of CPU/GPU tables.

\subsubsection{Indexed VS Operator}
\label{sec:indexed-operator}

The indexed operator uses a FAISS or cuVS index built on the embeddings of base tables (e.g., Reviews or Images). 
MaxVec supports two modes of vector search indices. 
In the first mode, the embedding vectors are stored within the index structure itself, which to our knowledge is the approach used by all current DBMS+VS engines. In the second mode, the embedding vectors are decoupled from the index structure and stored separately.
We adopt the terms data-owning and non-data-owning vector search indices to refer to these data structures.

\noindent\textbf{Data-owning index.}
The layout of embedding data within the index (row-wise, interleaved, etc.) is chosen by the index. The index also holds an ID in order for MaxVec to have a mapping back to the relational table row. The benefit of such a design is the flexibility to customize the layout for efficient distance computation, with separate optimizations across the CPU and GPU. The downside is a larger index, since the embeddings are part of the index, making it more expensive to move.

\noindent\textbf{Non-data-owning index.}
The index holds only the search data structure, the \texttt{embeddings\_type} data remain on the base table and are accessible during search. First, this design eliminates the redundant copy of the embeddings: much like a secondary B+ tree index, the index points into the base table rather than storing the data itself. Since FAISS does not natively support this layout, we patched it to decouple the index structure from the embedding storage. Second, compared to the data-owning design, when the index resides in CPU memory and the vector search operator is scheduled for GPU execution, only the compact search structure (centroids for IVF~\cite{sivic-ivf}, the graph for CAGRA~\cite{ootomo2024cagra}) must be transferred before execution begins, which is significantly less data than the full embedding set. Then, the GPU fetches only the necessary embeddings from host memory on demand during vector search. We cover how we implement the data transfer in the next section.

\noindent\textbf{Vector Index Data Movement optimizations.}
\label{sec:engine-optimizations}
Moving the index from CPU to GPU is decomposed into i) host to device (HtoD) transfers, ii) per-call HtoD setup, and iii) CPU index to GPU index transformation. To optimize the data movement cost, we implement three optimizations. We describe each mechanism here using index types and interconnects from those experiments as examples, and defer measurements to Sections~\ref{sec:transfer-time-optimizations} and~\ref{sec:non-owning}.
  
\textit{Caching} targets component (iii), the CPU index to GPU index transformation. MaxVec transforms the CPU index and caches it so that it can be directly used by the GPU constructor. Without caching, the index would be transformed during the critical path of the index movement for each query. One example of the cached transformations, is embedding layout conversion from a row-major format on the CPU to an interleaved format on the GPU. Another, for graph indices, is the HNSW$\to$CAGRA conversion, required because the index types differ in structure: HNSW (CPU)~\cite{malkov2018hnsw} is a multi-layer graph where the bottom layer (level 0) contains every node and its neighbor list, while CAGRA (GPU)~\cite{ootomo2024cagra} uses a single flat kNN graph. On each transfer, FAISS extracts the HNSW level-0 neighbors into a dense matrix, and converts the ID type to what the cuVS CAGRA constructor expects.

\textit{Pinning} targets component (i), the HtoD transfer time, by removing the host-to-host staging copy that pageable allocations require. Pageable host pages can be swapped by the OS, so it is not safe for the GPU DMA engine to address directly. When a \texttt{cudaMemcpy} is issued, the CUDA driver first copies the bytes into a temporary page-locked staging buffer in host RAM, then DMAs from there to the device~\cite{nvidia_cuda_pinned_memory}. Pinning skips this staging copy. Pinned memory is a scarce system resource, so MaxVec pins data selectively, for example index embeddings and base tables that repeatedly cross to the GPU. The pinning optimization is used on PCIe, as NVLink-C2C delivers the same bandwidth on pageable and pinned allocations~\cite{fusco2024understanding}.

\textit{Host-residency} targets components (i) and (ii). This optimization makes the non-data-owning design efficient by using hardware support. The unified-memory mechanism enabling this access is hardware-dependent. On GH200 (Table~\ref{tab:hardware}), the CPU and GPU sit on separate physical memory pools connected by NVLink-C2C. The GPU's MMU resolves host virtual addresses through the system page table via Address Translation Services (ATS), so the GPU reads host memory directly~\cite{nvidia_cuda_understanding_memory}. For partition-based indices, the per-partition \texttt{cudaMemcpy} calls used to copy embedding data are also eliminated, reducing the per-call HtoD setup count (ii).

\section{Experimental Analysis}
\label{sec:experiments}
Using the Vec-H benchmark and the MaxVec engine, we present an empirical analysis of SQL+VS queries on different engines, hardware, interconnects, execution strategies, index types and designs. 

\subsection{Experimental Setup}
\label{sec:exp-setup}

\noindent\textbf{Hardware \& Software}
We run three hardware configurations (Table~\ref{tab:hardware}) spanning PCIe~5.0, NVLink-C2C, and the DGX-Spark unified CPU-GPU memory system. The PCIe and NVLink machines have the same H100 GPU, isolating the interconnect's (I/C) effect. We use Apache Arrow v21.0.0~\cite{apache_arrow} for CPU, cuDF v25.08~\cite{cudf} for GPU, and Caliper~\cite{caliper} for instrumentation. For vector search, we use cuVS-enabled FAISS v1.13.0~\cite{douze2025faiss} linked against libcuvs v25.08~\cite{cuvs}. Builds use CUDA~12.8, GCC~13, and Ubuntu~24.04, except for DGX Spark (CUDA~13.0). We use PostgreSQL~17 with pgvector~v0.8.2~\cite{pgvector2025}. 

\begin{table}[t]
\caption{Hardware configurations.}
\label{tab:hardware}
\small
\resizebox{\columnwidth}{!}{%
\begin{tabular}{@{}llll@{}}
\toprule
 & \textbf{H100-PCIe~5.0} & \textbf{GH200-NVLink} & \textbf{DGX Spark} \\
\midrule
CPU & AMD EPYC 9124 & ARM Neoverse V2 & 10$\times$X925 + 10$\times$A725 \\
CPU cores & 16 & 72  & 20 \\
GPU & H100 NVL 94\,GB & H100 96\,GB & GB10 Blackwell \\
Mem.\ BW & 4\,TB/s (HBM3) & 4\,TB/s (HBM3) & 273\,GB/s (LPDDR5x) \\
Memory & 94\,GB HBM3 & 96\,GB HBM3 & 128\,GB unified \\
Interconnect & PCIe~5.0 & NVLink-C2C & --- \\
IC BW (1-way) & 64\,GB/s & 450\,GB/s & --- \\
\bottomrule
\end{tabular}
}
\end{table}

\noindent\textbf{Dataset \& Workload}
We use Vec-H with SF=1 where the dataset has $200$K parts, ${\sim}2.4$M reviews, and ${\sim}750$K images. We embed reviews with Qwen 0.6B~\cite{qwen3embedding} ($d{=}1024$, ${\sim}9.8$\,GB of embeddings) and images with SigLIP2~\cite{siglip2} ($d{=}1152$, ${\sim}3.4$\,GB of embeddings), giving a Rel:VS data ratio of $\sim\!1{:}12$. 
In the figures we use suffix $-r$ if the query runs vector search over reviews, and $-i$ if it is over images. 

\noindent\textbf{Engines.} PostgreSQL with pgvector is used as the baseline as it supports all Vec-H analytical SQL+VS queries across multiple index types. To evaluate \textit{gpu} and \textit{hybrid} (CPU/GPU) execution, we use MaxVec (Section~\ref{sec:maxvec_vector_operators}). PostgreSQL serves as a point of comparison for MaxVec's CPU implementation. In all cases, we tune PostgreSQL's configuration to use all available resources. For all engines, we load all data and indexes into memory before query execution. Total runtime comes from normal execution runs, while separate \texttt{EXPLAIN ANALYZE} or profiling runs are used to get additional information (i.e. the relational-to-VS operator time ratio).

\begin{table}[t]
\caption{Execution strategies for hybrid SQL+VS queries.}
\label{tab:exec_strategies}
\small
\resizebox{\columnwidth}{!}{%
\begin{tabular}{@{}l cc ccc ccc@{}}
\toprule
& \multicolumn{2}{c}{\textbf{Execution}} & \multicolumn{3}{c}{\textbf{Data on}} & \multicolumn{3}{c}{\textbf{Transfer}} \\
\cmidrule(lr){2-3} \cmidrule(lr){4-6} \cmidrule(lr){7-9}
\textbf{Strategy} & \textbf{VS} & \textbf{Rel} & \textbf{Idx} & \textbf{Emb} & \textbf{Rel} & \textbf{Idx} & \textbf{Emb} & \textbf{Rel} \\
\midrule
\textit{cpu}     & CPU & CPU & H & H & H & ---           & ---     & --- \\
\textit{gpu}     & GPU & GPU & D & D & D & ---           & ---     & --- \\
\textit{hybrid}  & CPU & GPU & H & H & H & ---           & ---     & copy \\
\textit{copy-di} & GPU & GPU & H & H & H & copy          & copy    & copy \\
\textit{copy-i}  & GPU & GPU & H & H & H & copy & stream  & copy \\
\textit{gpu-i}   & GPU & GPU & D & H & H & ---           & stream  & copy \\
\bottomrule
\end{tabular}
}
{\footnotesize \textbf{H} = host, \textbf{D} = device, {\textbf{copy}}: before execution, {\textbf{stream}}: read host memory at runtime}
\end{table}

\noindent\textbf{Measurement Methodology} 
We tune our query execution to achieve at least 95\% recall, and report query latency. In all cases and systems, each experiment is repeated 20 times, with the first repetition discarded as a warm-up. 
Unless otherwise noted, all host-side data (tables, index buffers, intermediate results) resides in non-pinned (pageable) memory. For all vector search queries we use $k{=}100$. Where necessary, we set the post-filter~\cite{wei2020analyticdb} oversampling $k'$ to $10\times k$, except Q15 which requires up to $500\times k$. This high oversampling exceeds the GPU top-$k$ cap enforced by FAISS (Section~\ref{sec:vech-quality}), so Q15 on the GPU has $<$50\% recall. We therefore explicitly analyze Q15 only in the CPU cases that meet the target recall, and report the GPU-achieved performance as a reference.

\noindent\textbf{Index types.} We cover the two dominant ANN types: partitioned (IVF via FAISS) and graph-based (HNSW on \textit{cpu} via FAISS, CAGRA on \textit{gpu} via FAISS/cuVS). The number suffix (e.g.\ IVF\textbf{1024}) is the partition count. Pgvector has both IVF and HNSW. We use \emph{Exhaustive} and \emph{ENN} interchangeably to indicate exhaustive vector search. The approximate \emph{ANN} cases are labeled by index type.

\begin{figure*}[t]
  \centering
  \includegraphics[width=\linewidth]{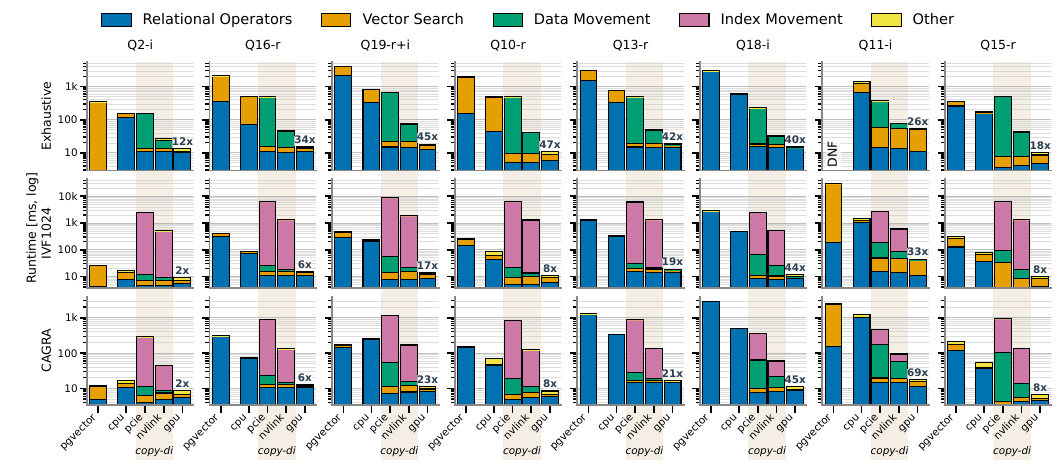}
  \caption{Vec-H per-query runtime with owning indexes (\emph{copy-di}) and baselines. All bars run on GH200 except \emph{pcie}, which runs on the H100 system. \emph{cpu} and \emph{pgvector} execute on the GH200 CPU, \emph{gpu} and \emph{nvlink} on the GH200 GPU.}
  \label{fig:plot1_current_indexes}
\end{figure*}

\noindent\textbf{Execution strategies.}  Table~\ref{tab:exec_strategies} sorts the execution strategies evaluated by where each vector search (VS) or relational (Rel) operator is executed, where the index, embeddings, and relational tables reside at query start (host H or device D), and what crosses the interconnect at run time. \textit{cpu} is the strategy used in current RDBMS+VS systems: all data resides and query is executed on the host CPU. \textit{hybrid} runs VS on the CPU and the relational side on the GPU, copying only the relational data to the device. The remaining four strategies run both operators on the GPU. In \textit{gpu}, all data is pre-resident on the device. \textit{copy-di} uses a \emph{data-owning} index that copies both embedding data (\emph{d}) and index structure (\emph{i}) per query, along with the relational tables. To the best of our knowledge, data-owning indexes are the default in every DBMS with integrated vector search. 
\textit{copy-i} and \textit{gpu-i} use a \emph{non-owning} index: only the index structure is copied to or kept on the GPU, while embedding rows stream from the host on demand, as the VS operator accesses them. \textit{copy-i} copies the index per query, \textit{gpu-i} keeps it resident on GPU memory.

\noindent\textbf{Time breakdown.} We report runtime for each configuration in bar plots, queries are ordered following their Vec-H taxonomy (Section~\ref{sec:benchmark}): VS@Start (Q2, Q16, Q19), VS@Mid (Q10, Q13, Q18), VS@End (Q11, Q15). Unless stated otherwise, the label on the rightmost (\textit{gpu}) bar of each per-query plot is the \textit{per-system CPU$\to$GPU speedup} for the MaxVec engine.
Each bar sums four components plus a small residual, which includes system overheads not belonging to a specific operator or transfer operation. \emph{Relational Operators} and \emph{Vector Search} represent the time to run their respective operators. \emph{Data Movement} covers CPU$\leftrightarrow$GPU transfers of relational tables, \emph{Index Movement} covers the transfer of the vector index structure (IVF/CAGRA) plus the embedding data if the index is data-owning (\emph{copy-di}). For ENN, the movement of embeddings is counted in the data movement. 

\subsection{State-of-the-art Strategies on CPU and GPU}
\label{sec:state-of-the-art}

Figure~\ref{fig:plot1_current_indexes} compares runtime of Vec-H queries when run on pgvector (cpu) and on MaxVec (cpu/gpu). MaxVec using \textit{cpu} runs Vec-H $1.6\times$--$20\times$ faster than pgvector on most (query, index) pairs. Comparing three vector search execution methods ENN, ANN with IVF, and ANN with graph (HNSW and CAGRA) indices, pgvector is faster in two cases (Q2, Q19 cagra), where its relational operators are faster.

The vector search is faster in MaxVec in every case: $3.5\times$--$8\times$ with ENN (Q11 on pgvector times out at $>\!20$-min), $1.6\times$--$7.1\times$ with ANN ($130\times$ on IVF1024 and $81\times$ on CAGRA for Q11). One reason for the general VS performance difference is that MaxVec's index and search implementation comes from FAISS, while pgvector implements its own. Another is the difference in vector operator design (Section~\ref{sec:maxvec_vector_operators}), for example Q11 achieves a much higher speedup than other queries because MaxVec's binary vector search operator takes the whole outer relation as a single query batch (Section~\ref{sec:maxvec_vector_operators}), while pgvector expresses Q11 with a \texttt{LATERAL} join (Section~\ref{sec:benchmark}) that performs one vector search operator call per each outer table tuple. \textbf{MaxVec is consistently faster than pgvector and thus we use it as our CPU baseline for the rest of the experiments section.}

\subsubsection{CPU vs GPU execution}
\label{sec:exec-strategies}
With the \textit{gpu} strategy (Figure~\ref{fig:plot1_current_indexes}), the data and vector index are already in the GPU memory before query execution starts. The result shows that MaxVec on \textit{gpu} beats \textit{cpu} on every (index, query) pair by $2\times$--$69\times$. Exhaustive search queries (ENN) on \textit{gpu} are within $\sim\!\!1.2\times$--$1.6\times$ of CAGRA, showing that GPU parallelism reduces the ENN VS cost significantly.

Next, we look at the \textit{copy-di} strategy, where the data and indexes start at the CPU and are copied to the GPU. NVLink (\textit{nvlink}) or PCIe (\textit{pcie}) interconnect is used depending on the hardware configuration. For ENN, the \textit{copy-di} strategy using nvlink is $4\times$--$19\times$ faster than \textit{cpu}. When pcie is used for ENN, the performance is a tie with \textit{cpu} on most queries, with Q11, Q13, Q18 winning by $1.4\times$--$3.7\times$. \textit{These results show that using ENN on GPU is a viable strategy when a fast interconnect is available, and ENN can be more robust with respect to tuning (i.e. oversampling in post-filtering\cite{fvs_chronis_survey_vldb} and choosing the right number of partitions for IVF~\cite{crack_ivf} or ef\_search for HNSW to achieve the requested recall \cite{malkov2018hnsw})}.

On the other hand, ANN with \textit{copy-di} is not competitive: IVF1024 and CAGRA on pcie are orders of magnitude slower than \textit{cpu}. This shows that data movement in data-owning indexes from host to device is a bottleneck. A faster interconnect, like nvlink, narrows but does not close the gap. Vector index transfer dominates \textit{copy-di} on ANN, taking over $95\%$ of wall time on most queries (${\sim}85\%$ on Q11). Note that the $index\_cpu\_to\_gpu()$ \footnote{\url{https://faiss.ai/cpp_api/file/GpuCloner_8h.html}} mechanism used in FAISS, captured by the ``Index Movement'' component of the bars, is far from achieving the theoretical bandwidth of pcie and nvlink. We analyze this further in Section~\ref{sec:index-transfer}.

\keyinsight{With current data-owning vector indexes, executing vector search on a GPU does not pay off, even with fast interconnects, unless the vector index is GPU resident}

\subsubsection{Which operators does the GPU accelerate more?}
\label{sec:gpu-savings-share}

\begin{figure}[t]
  \centering
  \includegraphics[width=0.9\linewidth]{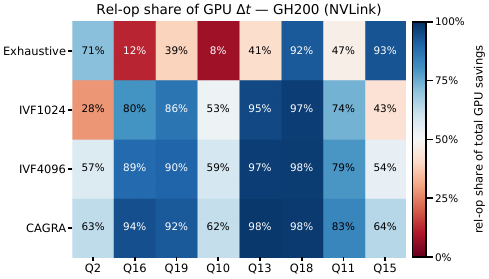}
  \caption{Share of total $cpu$ to $gpu$ wall-time savings attributable to relational operators.}
  \label{fig:plot2_gpu_savings_share}
\end{figure}

From Figure~\ref{fig:plot1_current_indexes}, we observed the \textit{gpu} execution strategy provides a clear benefit when the data is pre-resident on GPU. In Figure~\ref{fig:plot2_gpu_savings_share}, still on the GH200 system, we show the operator runtime savings of using the GPU compared to the CPU, with a fraction that captures the relational-operator acceleration:
$\mathrm{share}_{\mathrm{rel}} = (\mathrm{rel}_{\mathrm{CPU}} - \mathrm{rel}_{\mathrm{GPU}}) / (\mathrm{total}_{\mathrm{CPU}} - \mathrm{total}_{\mathrm{GPU}})$. Essentially, the larger the percentage, the bigger the share of the speedup that comes from accelerating the relational part of a query compared to the vector search. Above $50\%$, the benefit of the relational part is the majority.

The per vector search method median relational operator share of runtime improvement is $87\%$ for CAGRA, $77\%$ for IVF1024, $84\%$ for IVF4096, and $44\%$ for ENN. We observe two trends: First, for ENN, since \textit{cpu} execution is dominated by vector search, GPU benefits vector search more (Q16, Q19, Q10, Q13, Q11). Second, for ANN, even for the three queries (Q2, Q15, Q10) with the smallest relational operator improvement compared to VS, still the improvement of the relational operators is larger. This is a direct result of vector search being "cheaper" compared to relational operators. In this table, we include two configurations of IVF with 1024 and 4096 partitions, and we observe that the larger number of partitions makes VS faster and thus for the IVF4096 case, the relational operators improve relatively more compared to the IVF1024 case as the relational operator cost remains fixed.

\keyinsight{GPU execution yields a bigger improvement for relational operators compared to vector search.}

\subsection{Hybrid Execution}
\label{sec:pcie-hybrid}

\begin{figure*}[t]
  \centering
  \includegraphics[width=\linewidth]{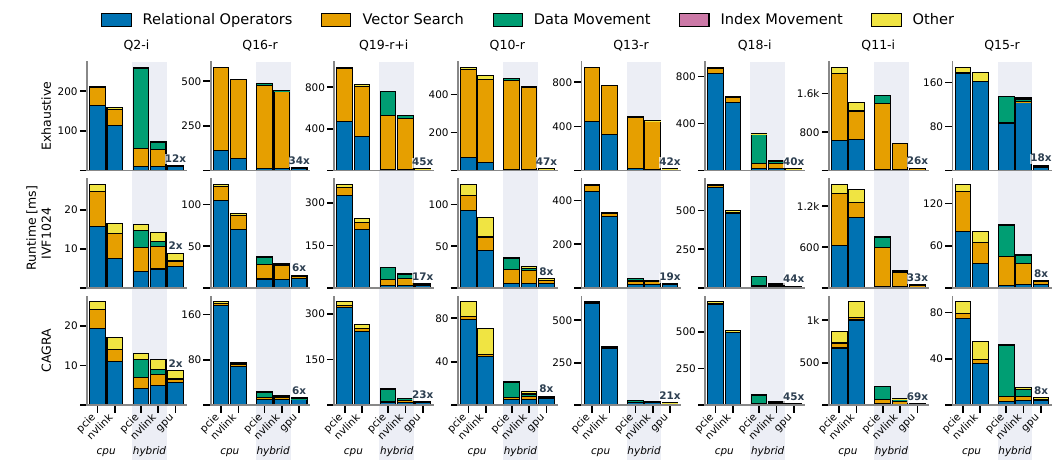}
  \caption{Vec-H per-query runtime under hybrid execution (VS on CPU, Rel on GPU). The \textit{pcie} \emph{cpu} and \textit{pcie} \emph{hybrid} run on the H100-PCIe system, while \emph{nvlink} \emph{cpu}, \emph{nvlink} \textit{hybrid}, and \emph{gpu}, on GH200.}
  \label{fig:plot3_hybrid}
\end{figure*}

The observations of Section~\ref{sec:state-of-the-art} motivate a \textit{hybrid} execution strategy: \textit{run relational operators on GPU and vector search on CPU.} 
The \textit{hybrid} strategy eliminates two large costs: the dominant index-transfer overhead of \textit{copy-di} (Figure~\ref{fig:plot1_current_indexes}), and the large amount of GPU memory needed to keep a device-resident index in \textit{gpu}. Figure~\ref{fig:plot3_hybrid} compares \textit{hybrid} to the \textit{cpu} and \textit{gpu} strategies. We report two CPU runtimes, one per platform: the AMD EPYC in the H100-PCIe system (pcie) and the ARM CPU in the GH200-NVLink system (nvlink). As the CPUs in the two hardware configurations are different, each \textit{hybrid} bar should be compared with the CPU bar from the same machine, e.g., pcie \textit{cpu} to pcie \textit{hybrid}.

\subsubsection{Where time goes in \textit{hybrid}}
\label{sec:hybrid-vs-rel-share}

In Figure~\ref{fig:plot3_hybrid}, we observe two trends. For most queries, when using ENN, vector search dominates the runtime. On the other hand, for most queries with ANN, the relational operators dominate. For ENN in \textit{hybrid} the time spent on VS is $90\%$+. The exceptions are Q2, and Q18 where the smaller embedding column makes VS on CPU cheap $\sim\!\!42$--$45$\,ms, and Q15, where the selective filter reduces the number of embeddings to be processed significantly. For \textit{ANN}, this relationship flips and relational operators dominate, even when executed on the GPU. On CAGRA \textit{cpu}-nvlink, where both run on the CPU, relational operators take $12\times$--$150\times$ longer than VS for $7$/$8$ queries (e.g.\ $495$\,ms vs $3.3$\,ms on Q18). Moving the relational part to the GPU in \textit{hybrid}-nvlink significantly improves runtime, reducing the gap to $1.2\times$--$4.8\times$.

\keyinsight{In analytical SQL+VS queries, relational operators can be orders of magnitude more expensive than vector search.}

\subsubsection{Hybrid Execution vs \textit{cpu} vs \textit{gpu}}
\label{sec:hybrid-strat}

Hybrid execution beats \textit{cpu} on $47$ of $48$ (query, index) pairs across pcie and gh200 machines.  
The only case where \textit{cpu} wins is ENN Q2 on pcie.
Compared to the \textit{gpu} strategy, \textit{hybrid} is slower but has significantly closed the performance gap. 
For example \textit{hybrid} using nvlink under CAGRA is within {$1.3$--$4.4\times$}, and IVF within {$1.6$--$5.7\times$} of the \textit{gpu} strategy. 

\keyinsight{With hybrid execution and ANN vector search, we get the best of both worlds: no data permanently residing on the GPU, GPU-acceleration for relational operators, and vector search acceleration through ANN indexes. }

Still, \textit{hybrid} is not optimal as the GPU is not used for VS, and depending on the query, intermediate results might have to be moved from the GPU to the CPU.


\subsection{Index Transfer Costs and Optimizations}
\label{sec:index-transfer}
\label{sec:owning-index-transfer-breakdown}

\begin{table}[t]
    \caption{Transfer times on reviews-table indexes (in ms).}
    \label{tab:cuda_transfer_stats}
    \centering
    \resizebox{\columnwidth}{!}{%
        \begin{tabular}{l rr rr | rr | r r}
            \toprule
            & \multicolumn{4}{c|}{\textbf{PCIe 5.0}} & \multicolumn{2}{c|}{\textbf{NVLink-C2C}} & & \\
            \textbf{Index}
                & \textbf{Total\phantom{\textsuperscript{P}}} & \textbf{HtoD\phantom{\textsuperscript{P}}}
                & \textbf{Total\textsuperscript{P}} & \textbf{HtoD\textsuperscript{P}}
                & \textbf{Total\phantom{\textsuperscript{P}}} & \textbf{HtoD\phantom{\textsuperscript{P}}}
                & \textbf{\#cpy} & \textbf{size (GB)} \\
            \midrule
            Flat/ENN               & 401 & 395 & 176 & 171 & 28.7 & 23.5 & 1 & 9.81 \\
            IVF1024            & {6510} & 452 & {6156} & 410 & {1266} & 46.4 & {5121} & 9.9 \\
            IVF4096            & {7890} & 349 & {7950} & 350 & {1423} & 62 & {20481} & 10.11 \\
            CAGRA             & 853 & 423 & 578 & 184 & 112 & 24.8 & 2 & 10.13 \\
            \midrule
            IVF1024\textsuperscript{H} & - & - & - & - & 4 & 2.5 & 3073 & {0.004} \\
            IVF4096\textsuperscript{H}  & - & - & - & - & 14.5 & 9.9 & 12289 & {0.017} \\
            CAGRA\textsuperscript{C} & 425 & 423 & 187 & 184 & 27.4 & 24.8 & 2 & 10.13 \\
            CAGRA\textsuperscript{C+H} & - & - & - & - & 0.8 & 0.76 & 1 & {0.307} \\
            \bottomrule
        \end{tabular}
    }
    {\footnotesize \textbf{P}: pinned memory, \textbf{C}: cached graph, \textbf{H}: host-resident (via ATS). \textbf{\#cpy}: total transfers}
\end{table}

The index movement throughput observed in Section~\ref{sec:state-of-the-art} fell well below the interconnect's peak bandwidth.
We now break down where the time goes during index movement and evaluate the optimizations MaxVec employs to minimize the cost of index movement (Section~\ref{sec:engine-optimizations}). 

Table~\ref{tab:cuda_transfer_stats} reports per-index transfer time on both interconnects: on NVLink, moving $\sim\!\!10$\,GB of embeddings ({ENN}) takes $28.7$\,ms while a similarly sized {IVF1024} index takes $1266$\,ms, a $44\times$ gap on identical data volume.
\textit{Total} is the wall-clock of FAISS \texttt{index\_cpu\_to\_gpu()}. 
\textit{HtoD} is the time spent in \texttt{cudaMemcpy} calls. The difference \textit{Total} $-$ \textit{HtoD} is per copy call setup and CPU to GPU index transformation. To avoid profiler interference, \textit{Total} is taken from an unprofiled run and \textit{HtoD} from Nsight Systems~\cite{nvidia_nsight_systems}. The top half of the table lists data-owning indexes (Section~\ref{sec:state-of-the-art}). The bottom half adds the optimized variants ($P$, $C$, $H$) that we evaluate next.

\noindent\emph{Pure HtoD bandwidth is not the bottleneck}. HtoD when issuing a single \textit{cudaMemcpy} call on a flat contiguous array of embeddings (Flat/ENN) reaches $\sim\!24$\,GB/s on non-pinned PCIe 5.0 and $\sim\!417$\,GB/s on NVLink-C2C. CAGRA's HtoD does the same. This is the expected interconnect bandwidth for pageable memory~\cite{maxbench_marko}.

\noindent\emph{Index movement is dominated by overhead}. FAISS has not optimized the transfer.
For CAGRA, \textit{total} is $\sim2\times$ HtoD on PCIe and $\sim4.5\times$ on NVLink, mainly because of the HNSW$\to$CAGRA conversion (Section~\ref{sec:engine-optimizations}).
For IVF, transfer overhead is extreme. Embeddings are stored in many per-partition arrays, transferred through small memcpys instead of a single large HtoD transfer. Surprisingly, IVF1024 issues $5{,}121$ \texttt{cudaMemcpy} calls and IVF4096 issues $20{,}481$, roughly $5$ per partition with mean size of $\sim\!1.8$\,MiB on IVF1024 and $\sim\!0.5$\,MiB on IVF4096. The effective throughput is $<2\%$ of the theoretical maximum bandwidth. \textbf{For index movement: data preparation and unoptimized data layout caps transfer throughput far below hardware peak performance.}

\noindent\textit{Pinning the memory and caching GPU indexes}
\label{sec:transfer-time-optimizations} Pinning ($P$) and graph caching ($C$) were introduced in Section~\ref{sec:engine-optimizations}. They improve the HtoD bandwidth and avoid transformation overhead (Table~\ref{tab:cuda_transfer_stats}). With ($P$), ENN and CAGRA HtoD rise from $\sim\!24$ GB/s to near peak $\sim\!55$ GB/s on PCIe 5.0. But, the transfer bandwidth for IVF does not improve, the copies are too small for pinning to matter and per-call HtoD setup dominates instead. With ($C$), CAGRA's \textit{Total} index transfer time improves from $853\!\to\!425$\,ms with unpinned memory in PCIe. With both $C$ and $P$, CAGRA is within $\sim\!2$ ms of \textit{HtoD} and \textit{Total} improves ($578\!\to\!187$\,ms pinned PCIe, $112\!\to\!27.4$\,ms NVLink).

\subsection{MaxVec Non-Owning Vector Indexes}
\label{sec:non-owning}

\begin{figure*}[t]
  \centering
  \includegraphics[width=\linewidth]{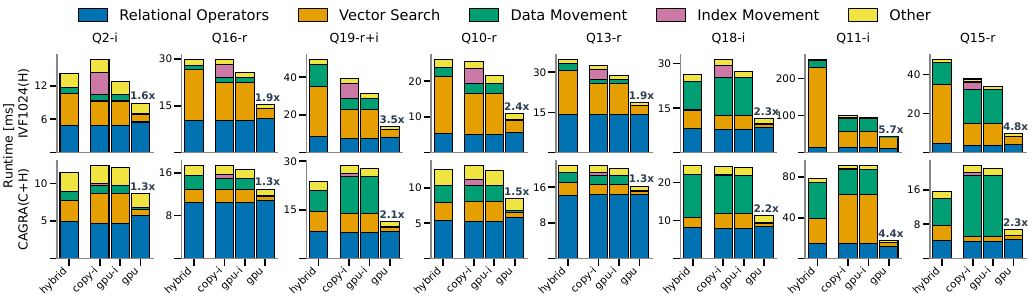}
    \caption{Vec-H per-query runtime on GH200-NVLink under the four optimized execution strategies (\emph{hybrid}, \emph{copy-i}, \emph{gpu-i}, \emph{gpu}). \emph{hybrid} runs VS on the CPU, the others on the GPU. Speedups annotated relative to \emph{hybrid}.}
  \label{fig:plot4_nvlink_nonowning}
\end{figure*}

In this section we evaluate the performance of non-data-owning indices for SQL+VS queries with the \textit{Host-residency} optimization, which is enabled by ATS over the GH200's coherent NVLink-C2C interconnect, a hardware capability not available in our PCIe machine (Section~\ref{sec:maxvec_vector_operators}). 

First, in Table~\ref{tab:cuda_transfer_stats} (bottom) we measure the transfer costs of the non-data-owning indices' metadata ($H$). Specifically, we measure the time to move the index metadata (centroids for IVF, graph for CAGRA), which takes place at the start of every query. Compared to moving both the embeddings and index metadata, like data-owning indices, the index movement reduces from $1266$\,ms to $4$\,ms ($316\times$) for IVF1024\textsuperscript{H}, and from $112$\,ms to $0.8$\,ms ($140\times$) for CAGRA\textsuperscript{C+H}.

We now evaluate the performance of full SQL+VS queries using a non-data-owning index with the \textit{copy-i} and \textit{gpu-i} execution strategies. \textit{copy-i} moves the index structure at the start of every query, \textit{gpu-i} takes advantage of the small size of the non-owning index structure ($4$\,MB IVF1024, $307$\,MB CAGRA) and keeps it resident in GPU memory prior to query execution. Both strategies only move the embeddings the ANN search visits to the GPU during query execution. In Figure~\ref{fig:plot4_nvlink_nonowning}, we see that compared to the \textit{gpu} strategy, both \textit{gpu-i} and \textit{copy-i} {are $\sim\!1.2-5.2\times$ slower than \textit{gpu}, while \textit{cpu} is $2-69\times$ slower than \textit{gpu} (Figure~\ref{fig:plot1_current_indexes})}. \textit{gpu} is faster but the margin {has improved}. Compared to the \textit{cpu} baseline on the GH200 system (\textit{cpu-nvlink} from Figure~\ref{fig:plot3_hybrid}), both \textit{copy-i} and \textit{gpu-i} are faster by $1$--$21\times$. We compare these non-owning strategies to \textit{hybrid} in Section~\ref{sec:choosing-strategy}.
  
\keyinsight{By updating vector index design to non-owning, and harnessing modern unified-memory hardware, we can accelerate full SQL+VS queries on a GPU without having to store
large data in GPU memory.} 

\subsection{Choosing a Strategy}
\label{sec:choosing-strategy}

\begin{figure}[b]
  \centering
  \includegraphics[width=1\linewidth]{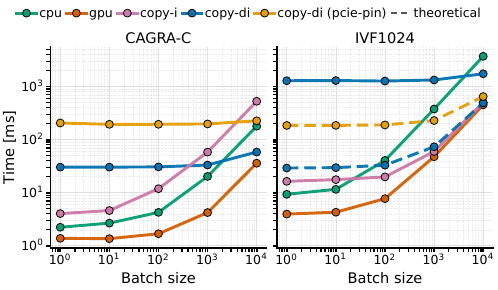}
  \caption{Vector search operator runtime on the reviews-table index, as we vary the input query vectors batch size on GH200-NVLink. Dashed \emph{theoretical} on IVF1024 assumes a single contiguous HtoD transfer.}
  \label{fig:plot7_crossover}
\end{figure}

Sections~\ref{sec:state-of-the-art} through \ref{sec:non-owning} evaluated six execution strategies for analytical SQL+VS queries: \textit{cpu}, \textit{gpu}, \textit{copy-di}, \textit{copy-i}, \textit{gpu-i}, and \textit{hybrid}. \textit{gpu} is the fastest in every case, but it requires the embeddings, the relational data and the index to be pre-resident in GPU memory, which is scarce and expensive. Three alternatives enable GPU acceleration without GPU-resident embeddings or large index movement overhead: \textit{hybrid} runs VS on the CPU and relational operators on the GPU. \textit{copy-i} runs both on the GPU with the index shipped per query, and \textit{gpu-i} runs both on the GPU with the index pre-resident. 

Looking at the optimized strategies for SQL+VS in Figure~\ref{fig:plot4_nvlink_nonowning}, there is no clear winner across all queries between \textit{hybrid}, \textit{copy-i}, and \textit{gpu-i}. \textit{gpu-i} beats \textit{hybrid} on $7/8$ IVF1024\textsuperscript{H} queries but only $4/8$ CAGRA\textsuperscript{C+H} queries, where the differences in runtime are within $5\%$. \textit{copy-i} is roughly even with \textit{hybrid} on IVF1024\textsuperscript{H} and slightly behind on CAGRA\textsuperscript{C+H}. Q11, the batched deduplication query, is the strongest case for VS-on-GPU under IVF, where both \textit{copy-i} and \textit{gpu-i} run $\sim\!\!2.5\times$ faster than \textit{hybrid}. The IVF-vs-CAGRA gap suggests IVF benefits more from GPU acceleration than CAGRA. 

\subsubsection{Which method should you choose?}
\label{sec:what-to-do}

Based on all our empirical observations, we derive a decision heuristic to pick the best execution strategy depending on GPU memory availability and the type of vector index: 

\textit{Choose {gpu} when all data and indices fit on the GPU. When only the index fits, choose {gpu-i} for IVF and {hybrid} for CAGRA. When neither fits, {hybrid} is best, with {copy-i} as an alternative for IVF at large vector query batch sizes.}

The fastest method between \textit{hybrid}, \textit{copy-i}, and \textit{gpu-i} can depend on many parameters: index type, batch size, scale factor, target recall, host hardware.
To illustrate, we isolate and vary the query batch size on a pure VS micro-benchmark, tuned for $90\%$ recall. For a DBMS with VS, a large batch size query is equivalent to a vector similarity join~\cite{fast_annjoin_paper}, which matches the Q11 VS pattern in our workload (here without the relational operators). We ask at what batch size does the GPU VS speedup amortize the cost of moving just the index (\textit{copy-i}), and does it ever become worth it to have a data-owning index, which also moves the embeddings to GPU memory (\textit{copy-di}). 

For IVF1024 (Figure~\ref{fig:plot7_crossover}, right) the index movement cost is amortized quickly. \textit{copy-i} becomes better than \textit{cpu} between $10$--$100$ queries and converges to \textit{gpu} performance at large batch sizes. For CAGRA (Figure~\ref{fig:plot7_crossover}, left), moving just the index structure is not enough, \textit{copy-i} loses to \textit{cpu} at every batch size, and only \textit{copy-di} improves over \textit{cpu} at above a $10^3$-query batch.

Across our SQL+VS experiments, non-data-owning indexes make GPU execution the default for both VS and relational operators. The index structure is small enough to fit or be copied to the GPU and only ANN visited embeddings stream from host during search. \textit{Hybrid} remains the recommendation for HNSW/CAGRA and when only the relational operators need to be accelerated. As DBMSs with VS support mature, the query optimizer will estimate data movement, operator runtimes, batch size, GPU memory budget, and recall target to identify the optimal per-query execution strategy.


\subsection{Does the GPU Still Win Without HBM?}
\label{sec:nonowning-spark}

\begin{figure*}[t]
  \centering
  \includegraphics[width=1\linewidth]{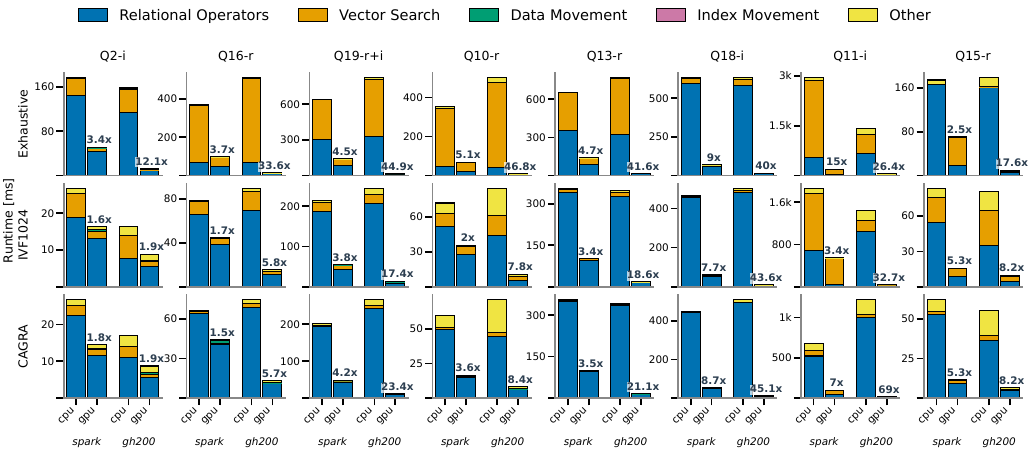}
  \caption{Per-query Vec-H runtime on DGX-Spark and GH200 for the \textit{cpu} and \textit{gpu} execution strategies.}
  \label{fig:plot8_spark}
\end{figure*}

Finally, we look at the GB10 Superchip on DGX-Spark~\cite{nvidia_dgx_spark} (Table~\ref{tab:hardware}), which puts a Blackwell GPU and a 20-core ARM CPU on a single package, sharing one $128$\,GB LPDDR5x pool. We look at this system for three reasons. First, it is a fraction of the cost of a GH200 system~\cite{nvidia_grace_hopper}, so we can ask whether the SQL+VS speedups we reported on the GH200 generalize to less expensive hardware. Second, the GPU on Spark has no faster memory tier than the CPU: both draw from the same LPDDR5x at $273$\,GB/s, ${\sim}15\times$ slower than the HBM on a GH200, so any GPU speedup must come from compute (parallelism, Tensor units) rather than from a memory-bandwidth advantage. Third, the host-versus-device memory boundary that drove Sections~\ref{sec:state-of-the-art} through~\ref{sec:non-owning} (where data resides, what crosses the interconnect, when to copy) does not exist. A SQL+VS query on this hardware can indistinctly run on the CPU or the GPU.

Figure~\ref{fig:plot8_spark} compares the \textit{cpu} and \textit{gpu} execution strategies on Spark. We can see that \textit{gpu} beats \textit{cpu} on every (query,\,index) pair, by $\sim\!2.5$--$15\times$ for ENN and $\sim\!1.5$--$9\times$ for ANN. 
The DGX-Spark, without the fast HBM bandwidth or the faster GPU is, as expected, slower in runtime compared to the GH200. The GH200 performs $\sim$4–8$\times$ faster for ENN and $\sim$1.5–13$\times$ faster for ANN on \textit{gpu}. 

The architecture and low cost make the DGX-Spark an attractive system for SQL and SQL+VS workloads. On the single-tier unified-memory hardware the placement question of Sections~\ref{sec:state-of-the-art} through \ref{sec:non-owning} no longer applies, and the \textit{gpu} strategy can always be used.

\keyinsight{Single-tier unified-memory systems like the DGX-Spark offer simpler data placement and GPU-acceleration at a fraction of the cost. Well-suited for database workloads and a possible direction for GPUs to better support relational engines.}

\section{Conclusion}
\label{sec:conclusion}

This work explores whether GPUs can accelerate analytical SQL+VS queries, where the cost lies between relational and vector search operators, and how vector indices and execution should be designed for heterogeneous CPU-GPU database systems. To answer the question, we developed Vec-H, an analytical SQL+VS benchmark extending TPC-H with two embedding tables and eight queries, and MaxVec, an execution engine for full SQL+VS queries on heterogeneous CPU-GPU hardware, with per-operator hardware placement and transparent data and index movement. We evaluated them across PCIe~5.0, NVLink-C2C, and DGX-Spark unified memory, varying execution strategies, index types, and index designs.

From this analysis, we find that GPU acceleration does pay off for analytical SQL+VS, but the win is not driven by faster vector search. Most of the benefit comes from accelerating the relational part of the query. With the data-owning indexes used in current DBMS+VS systems, GPU vector search itself does not pay off even on fast interconnects. However, with non-data-owning indexes and hybrid execution strategies, GPU acceleration can be enabled without having to keep the data GPU-resident. The right choice of execution strategy depends on the index, interconnect, and the GPU memory budget. To benefit from GPU acceleration in practice, as DBMS+VS systems mature they need query optimizers that pick the optimal per-query execution strategy. A final important insight is that in unified memory architectures like the DGX Spark, running both the relational and the vector search components on the GPU always wins, providing a development direction for appliances focused on data analytics.

\section{Acknowledgments}
\label{sec:acknowledgments}

Work supported by a grant from the Swiss AI initiative and Swiss National Supercomputing Centre (CSCS) under project ID sm94.

\bibliographystyle{ACM-Reference-Format}
\bibliography{bibliography}

\end{document}